\documentclass[]{article}
\usepackage{multirow}
\usepackage{hyperref}
\usepackage{setspace}
\usepackage{multirow}
\usepackage{hyperref}
\usepackage{amsmath}
\usepackage{enumerate}
\usepackage{graphicx}
\usepackage[margin=2.5cm]{geometry}
\usepackage{caption}
\usepackage{epsfig}
\usepackage{parskip}
\usepackage{rotating}
\usepackage{pdflscape}
\usepackage{verbatim}
\usepackage{float}
\restylefloat{table}

\begin{document}

\title{Univariate and data-depth based multivariate control charts using trimmed mean and winsorized standard deviation}
\author{Kushal Kr. Dey$^{+}$, Kumaresh Dhara$^{+}$, Bikram Karmakar$^{+}$, Sukalyan Sengupta$^{*}$}
\date{}

\maketitle

$+$ Graduate student, Indian Statistical Institute, Kolkata\\[2 pt]
$*$  Professor, Statistical quality control and Operations Research Unit, Indian Statistical Institute\\[2 pt]
Email: \\ \href{mailto: kshldey@gmail.com}{\nolinkurl{kshldey@gmail.com}} \hspace{0.4 cm} \href{mailto: kumaresh.dhara@gmail.com}{\nolinkurl{kumaresh.dhara@gmail.com}} \hspace{0.4 cm} \\  \href{mailto: bikram.karmakar@yahoo.in} {\nolinkurl{bikram.karmakar@gmail.com}} \hspace{0.4 cm} \href{mailto: sukalyan@hotmail.com}{\nolinkurl{sukalyan@hotmail.com}}\\[1 pt]
Tel: +91-9432640899, +91-9433776884, +91-9007257984 \\[1 pt]
Address: 203, Barrackpur Trunk Road, Kolkata-7000108, India\\

\begin{abstract}
Over the years, the most popularly used control chart  for statistical process control has been Shewhart's $\bar{X}-S$  or $\bar{X}-R$  chart along with its multivariate generalizations. But, such control charts suffer from the lack of robustness. In this paper, we propose a modified and  improved version of Shewhart chart, based on trimmed mean and winsorized variance  that proves robust and more efficient. We have generalized this approach of ours with suitable modifications using depth functions for Multivariate control charts and EWMA charts as well. We have discussed the theoretical properties of our proposed statistics and have shown the efficiency of our methodology on univariate and multivariate simulated datasets. We have also compared our approach to the other popular  alternatives to Shewhart Chart and established the efficacy of our methodology. 
\end{abstract}


\textbf{Keywords}: Shewhart control charts, Hotelling $T^2$ charts, EWMA control charts, Depth functions, Trimmed Multivariate distribution, Bootstrapping


\textbf{AMS subject classification} 62P30 (Applications in engineering and industry); 62F40 (Bootstrap, Jackknife and Other resampling methods); 62F35 (Robustness and Adaptive Procedures); 62H11 (Directional data; Spatial Statistics)

\section{Introduction}

Shewhart (1931) introduced control charts in  the mid 19th century as  means for statistical process control [1]. Shewhart developed both the  $\bar{X}-R$ and  $\bar{X}-S$ control charts, but initially, owing to its ease of construction, $\bar{X}-R$ chart was more preferred. However, nowadays, with high speed production processes, large subgroups can be easily obtained. Therefore, $\bar{X}-S$ charts are more relevant because standard deviation is a better measure of dispersion than the range. Later on, a multivariate analogue of Shewhart chart was introduced on the basis of the Hotelling's $T^{2}$ statistic [3]. A major criticism of the Shewhart chart and its multivariate analogue entails from the fact that both mean ($\bar{X}$) and standard deviation ($S$)  are non-robust measures of central tendency and dispersion. Another drawback of $\bar{X}-S$ type of control charts is its poor performance under small mean shifts. Various charts have been proposed using robust measures  like runs test [7], repeated median filters [10] or univariate trimmed mean and range [16].  These methods though efficient, lack interpretability and are computationally difficult. Liu [1995] proposed a \emph{r chart} based on the ranks of the data points for multivariate data [20]. She asserted that the ranks, when scaled to [0,1] will follow a uniform distribution and hence, the empirical distribution of the ranks would converge in law to $U(0,1)$. A problem with this method is that it neglects the true values of the observations and considers only relative percentiles. As a result, even if an observation undergoes moderate change, the change may not be reflected in the \emph{ r chart}.
 In this paper we have proposed control charts based on trimmed mean and winsorized standard deviation and discussed its distributional features. We have extended our notion to  EWMA (\emph{Exponentially Weighted Moving Average}) control charts. We have also suggested alternatives based on our approach, to  Hotelling $T^2$ charts [3] and  Multivariate EWMA charts  using data depth. 

The distributions of the measures of central tendency and dispersion have been studied using re-sampling methods like \emph{Bootstrapping}. We have compared the $ARL$ ({average run length}) performance of our proposed chart with that of the usual mean and variance based control charts. The results of the study have established that in the presence of outliers, our proposed control chart clearly outshines the standardl charts, and has comparable performance with the latter in absence of outliers. Therefore the use of such charts is strongly recommended. All the relevant developments have been discussed in the subsequent sections. The programs have been written and evaluated in statistical softwares like  MATLAB2009a and R2.11.1.


\section{Definition of trimmed mean and winsorized s.d.}

Univariate trimmed mean was proposed by Tukey (1963) as a robust estimate of process average [27]. Let $x_1,x_2,...,x_n$ be a sample of size $n$ on measurement of a particular quality characteristic. Then the $100\alpha\%$ trimmed mean is defined as 
$$ \bar{X}_{t}=\frac{\sum_{i=t+1}^{n-t}x_i}{(n (1-2\alpha))} $$
where $\alpha \in (0,1)$; $t=[n\alpha+.4]$. For simplification, Iglewicz and Langenberg (1986) have taken $t$ to be the floor of $(n\alpha+.4)$ as an approximation [16]. The Tukey trimmed mean follows asymptotically normal distribution and its standard error is defined as 
$$se_t=s_w/\sqrt n (1-2\alpha)$$ 
where $s_w$ is the $100\alpha\%$ Winsorized standard deviation. However the distribution of $se_{t}$ is not known. By definition, this statistic is robust.

The multivariate definition of trimmed mean depends on the choice of an appropriate depth function.

$$\bar{X}_{t}=\frac{\sum_{depth(\tilde x_i)>cutvalue}{\tilde x_i}}{(\#(\tilde x_i,(depth(\tilde x_i)>cutvalue))}$$

where $depth(\tilde x_i)$ is the value of the chosen depth function for the $i^{th}$ vector observation  $\tilde x_i$ and \emph{cutvalue} is the minimum value of depth that would be accepted in data trimming. We further define Winsorized variance for multivariate observations to be $S_{w} = Disp((\tilde y_1,\tilde y_2,\ldots,\tilde y_n))$, where $Disp$ is the estimated dispersion matrix computed from the $\tilde y_i$'s where $\tilde y_i = \tilde x_i$ if $depth(\tilde x_i)>cutvalue$ and $\tilde y_i = x_i^\star$  otherwise where, $ x_i^\star$ is the observation $\tilde x_i$ with minimum depth above $cutvalue$.

The four types of depth functions used for trimming by us are,

\begin{enumerate}
\item Spatial depth or L-1 data depth (Chaudhuri, P. (1996))
\item Tukey depth (Tukey, J. W. (1975))
\item Liu depth or Simplicial depth (Liu, R. Y. (1988))
\item Oja depth (Oja, H. (1983))
\end{enumerate}

the reasons being  that they have desirable properties like affine invariance, maximality of center, monotonicity w.r.t. the deepest point and vanishing at
infinity.

\section{Theoretical Development}

First, we present a theoretical background corresponding to the various control charts that form the basis of our study.

\subsection{Modification of the Shewhart chart }

\begin{figure}\label{Figure1}
\centering
\includegraphics[width=15cm,height=9cm]{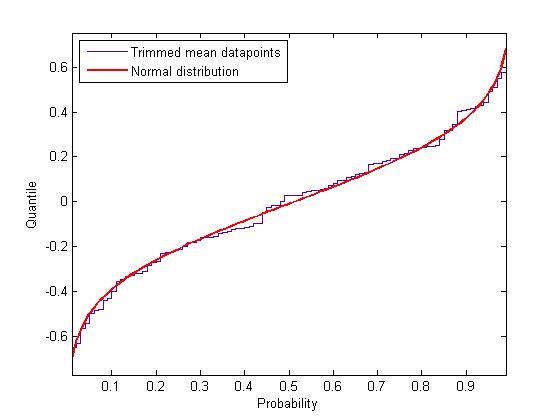}
 \caption{QQ plot of trimmed mean vs normal}
 \label{fig:normalqqplot}
\end{figure}

\begin{figure}\label{Figure2}
\centering
\includegraphics[width=15cm,height=9cm]{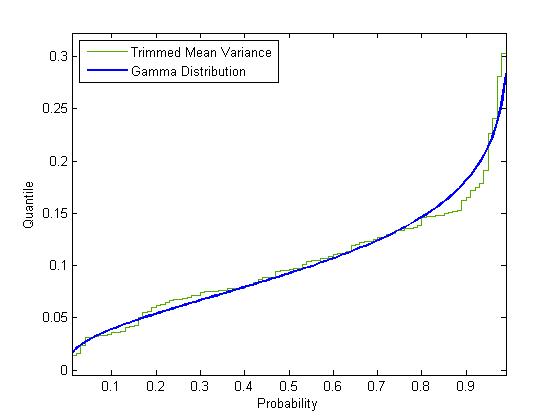}
\caption{QQ plot of trimmed mean variance vs gamma}
 \label{fig:gammaqqplot}
\end{figure}

In Shewhart control chart ($\bar{X}-S$), the distribution of $\bar{X}$ and S under the normality assumption was used in order to determine the central line and the control limits of the chart. If the data comes  from a normal distribution, the mean of the data will also follow normal distribution and the probability that the mean of the subgroups of size $n$ go beyond the $\pm3\sigma/\sqrt n$  limit is as small as 0.0023. We use $\bar{\bar{X}}\pm3S/\sqrt n$ as the control limits in the $\bar{X}$ control chart, the standard error of the subgroup means being an unbiased estimator of $\sigma$. The distribution of S under normality of subgroup means  was used for constructing the S-chart. We propose the use of  trimmed mean $\bar{X}_t$  and its standard error, $S_t$ in developing the charts. The control limits of the charts will follow from the distributional features of these two quantities. If the observations($x$) are from $N(\mu,\sigma^2)$ then one can take the transformation $Z=(x-\mu)/\sigma\sim N(0,1)$. Thus standard normal distribution is taken as the reference distribution for our discussion. It has been observed by us from simulation studies that trimmed mean($\bar{X}_{t}$) is asymptotically normal (Figure ~\ref{fig:normalqqplot}) and ${S_t}^2$ asymptotically follows a Gamma distribution whose parameters can be estimated from the data (Figure ~\ref{fig:gammaqqplot}).\\

\subsection{Modification of the EWMA chart}

The usual EWMA control chart building mechanism is as follows. Let $\bar{X}_{i}$ be the $i^{th}$ subgroup mean, then  we define an exponentially weighted average for each $i$ from 1 to $n$, $Z_{i}=\lambda\bar{X}_{i}+(1-\lambda)Z_{i-1}$, $\lambda$ being a constant where $Z_{0}=\mu_{0}$, the process target value, which we consider to be $\bar{\bar{X}}$, mean of subgroup means, in our study. It is known that for large number of subgroups, the in-control $ucl$ and $lcl$ may be chosen to be $\mu_{0} \pm L\sigma\sqrt{(\lambda)/(2-\lambda)}$ for suitable choice of $L$ and $\lambda$ respectively with the central line being $\mu_{0}$. For our proposed analogue, we replace the subgroup means by $100\alpha\%$ trimmed subgroup mean. The mechanism is same as in the case of Shewhart chart, though here we do not use any distributional features of $\bar{X}_t$ and $S_t$ for forming the control chart, instead we just plug in $\bar{\bar{X}}_{t}$ and $S_{t}$ as robust estimators of $\mu_{0}$ and $\sigma$ in the usual control limits. The choice of L and $\lambda$  though are subjective as in the usual case.

\subsection{Modification in Hotelling's $T^2$ chart }

Assume that we have a process at hand that generates multivariate observations, say $p$-variate observations $\tilde X  =$\newcommand{\vect}{(X_1,X_2,\ldots,X_p)}$\vect$ and we need to ensure that the process stays in control over time. One way to do it is to treat variables separately and construct control charts  for each of them. But that approach besides being laborious and time consuming, neglects the correlation among the variables. Hotelling's $T^2$ control charts are the most popular multivariate control charts in general. The Hotelling $T^2$ statistic, $ (\bar{X}-\mu)^{\mathrm{T}}{\Sigma}^{-1}(\bar{X}-\mu)$ follows ${{\chi}^2}_p$ under normality assumption, where $p$ is the number of variables, for a sample of size n. Once we estimate $\Sigma $ by \textbf{S}, the unbiased estimator then ${(n-p)/p}(\bar{X}-\mu)^{\mathrm{T}}{S}^{-1}(\bar{X}-\mu) \sim F_{p,n-p} $. \\
We suggest a modification to the ordinary Hotelling $T^2$ statistic in order to make it robust. The new test statistic proposed is 
$${\tau}^2 = (\bar{X}_{t}-\mu)^{\mathrm{T}}{\textbf{S}_w}^{-1}(\bar{X}_{t}-\mu).$$ 
where $\bar{X}_{t}$ and $S_{w}$ are as defined in \textbf{Sec:2}. It is very difficult to determine the distribution of  ${\tau}^2 $ accurately since it depends on various factors such as the  subgroup size($k$), choice of the depth function ($depth$), \emph{cutvalue} of trimming  etc. So we had to resort Bootstrap techniques.

\subsection{Modifications in MEWMA chart }

The univariate definition of EWMA chart easily extends to the multivariate case. For p-variate observations, define p-variate vector $Z_{i}=\Lambda\bar{X}_{i}+(1-\Lambda)Z_{i-1}$, where $\Lambda$ denotes a $p \times p$ matrix , $\bar{X}_{i}$ $"i"$th subgroup mean vector and $Z_{0}=\bar{\bar{X}}$. Assume $\Lambda$ to be a scalar matrix with diagonal element $\lambda$ for simplicity of calculations  and $\Sigma_{Z}={\frac{\lambda}{2-\lambda}}\Sigma$, where $\Sigma$ is the variance covariance matrix of subgroup means. We know that
$(Z_{i}-\mu_{z})^{\mathrm{T}}{\Sigma}_{z}^{-1}(Z_{i}-\mu_{z})$, for  mean $\mu_z$ and variance-covariance matrix $ {\Sigma}_{z}$ follows ${{\chi}^2}_p$ distribution under the normality assumption of $Z_{i}'s$. The quantity to be plotted in the MEWMA control chart is $ (Z_{i}-\hat{\mu_z})^{\mathrm{T}}{\hat{\Sigma}_z}^{-1}(Z_{i}-\hat{\mu_z})$ for each i , where $\hat{\mu_{z}}$ and $S_{z}$ are the sample counterparts of the mean and variance of the $Z$ values respectively. We, however propose to plot the statistic $Z_{{i}_{t}}=  \lambda\bar{X}_{{i}_{t}}+(1-\lambda)Z_{{i-1}_{t}}$  for $"i"$ th sample value of $Z$, where $\bar{X}_{{i}_{t}}$ is the trimmed mean due to the "i"th subgroup. We use \\
$$\psi^2 = (Z_{{i}_{t}}-\hat{\mu_z})^{\mathrm{T}}{{S}_{zw}}^{-1}(Z_{{i}_{t}}-\hat{\mu_{z}})$$ 
as our test statistic for the $i$th sample value of $Z$,where  $ {S}_{zw}$ is the sample winsorized variance-covariance matrix of the $Z$ values. But again it is difficult to find the distribution of our proposed statistic and we have to resort to Bootstrap techniques in order to construct the control chart. 

\section{Comparative simulation study}

\subsection{Adopted rules of Insertion of Outliers}

Trimmed mean control chart is expected to give better result than the mean and variance based control chart  when the process is in control but the data contains some outliers. In order to test the comparative performance of our chart, we need to insert certain outliers to our simulated data. We adopt the outlier insertion rules due to Gupta and Sengupta [10] in this regard.

\begin{itemize}
\item Each subgroup contains a fixed number of outliers.

\item Each outlier is actually a random number from a largely shifted distribution.
\
\item The place where the outlier is to put is done by randomly choosing the point.

\item  The sign of the outlier is taken +/- with probability half to each of them.
\end{itemize}

In our studies, for 0.25 shift and for 10\% trimming we took one outlier and for 20\%trimming we inserted two outliers in each subgroup.

\subsection{Univariate Case}

We carried out a number of simulation studies in order to test for the performance of the chart proposed by us compared to the standard mechanisms. We considered our reference distribution to be standard normal, because given a normal density with specified mean and variance, we can use the Z-transformation to bring it to standard normal and apply the same set of procedures as discussed above.  In Phase I, we simulated 80 rational subgroups of fixed subgroup size from the standard normal distribution ( in control distribution) and used these  to construct the control limits for both Shewhart's $\bar{X}-S$ and our proposed $\bar{X}_{t}$- $S_{t}$ charts. In Phase II,  10,000 subgroups of observations are generated from the reference distribution, subgroup means are plotted and  the $ARL$ is recorded. The subgroup sizes for the univariate Shewhart chart have been taken to be 10, 11, 12, 14, 15 and 20, because such  sample sizes are more used in practice in industrial processes. For EWMA chart,  we have taken  subgroups of size 20 and fixed our $L$ at a convenient level of 3 (standard choice). We have taken $\lambda$ to be 0.20,0.25 and 0.40, which are the preferred choices (Crowder and Stephen [9], Lucas \emph{et al.} [22], Hunter [15]) and $L$ to be 3.We have observed the performance of the EWMA chart corresponding to three values of $\lambda$ namely 0.4,0.25 and 0.2.The trimming percentage is usually taken to be 10$\%$ or 20$\%$ for each subgroup. The $ucl$ and $lcl$ have been taken to be $95^{th}$ and $5^{th}$ percentile points for the proposed charts. The control charts presented at the back of our report (Table ~\ref{table:arluniv}) and (Table ~\ref{table:arlewma}) give a comparison of our proposed chart with the Shewhart chart and EWMA chart in 3 scenarios-no shift, under small mean shift and in the presence of an outlier.


\subsection{Multivariate case}

\subsubsection{Hotelling's $T^2 chart$}
For multivariate case ,we do not standardize the data because of the lack of significance of it. We take the subgroup size to be 20. For trimming the data ,we have used our previously mentioned four depth functions and compared the control chart performance under each case. But the data trimming will obviously depend on the choice of the $cutvalue$ used for trimming. We simulate data from a bivariate normal distribution with $\mathbf{\mu}$=
$\begin{pmatrix}
0\\
0
\end{pmatrix}$ and variance covariance matrix $\mathbf{\Sigma}$=
$\begin{pmatrix}
1 & 0.3\\
0.3 & 1.2\\
\end{pmatrix}$ in 1000 subgroups of size 20 each. To get an optimal cut-off value we select a large sample say of 100 in control subgroups from the process and find the depths of all the points in each subgroup. For MEWMA chart, choice of $L$ and $\lambda$ are same as in univariate EWMA chart. Since the distribution of the test statistic of interest is not known, we use bootstrapping. 

We first drew 100 subgroups of observations  of same subgroup size when the process is in control.
 We selected 1000 subgroups of same size with replacement  from these 100 subgroups and computed  trimmed mean $(\bar{X}_{i_{t}})$ and winsorized dispersion matrix ${\textbf{S}_{i_w}}$ for $i$th subgroup for a given the depth function. We computed  the grand trimmed mean and mean of the subgroup  winsorized dispersion matrices,
$$\bar{\bar{X}}_{t} = \frac{\sum_{i=1}^{1000}{\bar{X}_{i_{t}}}}{1000}$$  $$\bar{\textbf{S}}_w=\frac{\sum_{i=1}^{1000}{\textbf{S}_{i_w}}}{1000}$$
For each subgroup we computed ${\tau}^2$ for each subgroup and calculated its empirical distribution.  

For MEWMA chart, a similar procedure of resampling is adopted. Here, we compute $Z_{i_{t}}=\lambda\bar{X}_{i_{t}}+(1-\lambda)Z_{{i-1}_{t}}$ for each $i$, taking $\bar{\bar{X}}_{t}$ to be $Z_0$. Then we compute $\bar{Z}_{t}$, the grand mean of all the $Z_{i_{t}}$ values. We estimate the dispersion matrix of the Z variable 
$$\textbf{S}_{Z}=\frac{\lambda}{1-\lambda}\bar{\textbf{S}}_w$$
1000 $\psi^2$ values are obtained and empirical distribution is computed. For distribution of both $\tau^{2}$ and  $\psi^2$, we choose $90$th percentile point as $ucl$ and 0 as $lcl$ as the process is out of control only when these statistics are significantly high.  We have compared our recommended multivariate control chart under various depth functions with the standard charts under no shift, small mean shifts and in presence of outliers.(Table ~\ref{table:arlmulti} and Table ~\ref{table:arlmultimulti}). 

Next, we have compared our approach with the two approaches due to Liu [20] based on ranks and another approach based on MCD estimators by Chenouri \emph{et al}[28]. We considered a samples of size 20 from $N(0,1)$ and tested for the comparative efficacy of the individual algorithms under mean shifts and presence of outliers. The results are reported in Table~\ref{table:liumcd}. Liu's method performed exceptionally well in case of mean shifts but was also very responsive to the presence of outliers, while MCD method, despite being well adapted to the presence of outliers, had poor performance under small mean shifts. Our method on the other hand had good performance both under mean shifts and presence of outliers. So, the strength of our algorithm as an alternative to standard charts is quite obvious. 

\section{Discussions}
\begin{itemize}
\item For a process in control or for small mean shifts,there is not much to choose between  the  mean based control chart and our proposed control chart. However, as expected, our recommended chart performs way better than the Shewhart and EWMA charts and their corresponding multivariate analogs, that too for any choice of depth function in multivariate case. 
\item We have preferred subgroup size around 20, because very small subgroup sizes often lead to unusual fluctuations in the $arl$ values. Some depth functions, like Tukey's depth are not at all reliable for small sample sizes and may cause excessive data loss on trimming. \\[1 pt]
\item It has been observed that bad choice of cut-off value often leads to highly fluctuating $ucl$ and $lcl$ values and the control chart no longer stays very reliable. That is why we have considered estimation of the $cutvalue$ in the multivariate case to serve this purpose. we have considered the distribution of the depth function corresponding to all the points and taken a quantile corresponding to the distribution as $cutvalue$ so that neither is there a huge data loss nor a complete retainment of data in most cases. \\[1 pt]
\item Some depth functions have limitations of application. Tukey's depth can only assume a limited range of values for subgroup sizes like 10 or 20. So, in such cases this depth is not at all reliable. Liu depth can be used for bivariate data only and cannot be extended to higher \\ dimensions . Spatial depth and the Oja depth have been very consistent in their performances under various scenarios viz-change in subgroup size,change in $\lambda$ values,dimensionality etc as evident from Table ~\ref{table:arlewma} and Table ~\ref{table:arlmulti}, so these are more preferable under general circumstances compared to Liu and Tukey depth. \\[1 pt]
\item From bootstrapping up to the process surveillance stage, subgroup size should not be changed a lot. That's because the $ucl$ and $lcl$ are obtained from an empirical distribution for a given sample size. So, with sample size, it is expected to alter as well. But it has been seen up to k$\pm2$ not much deviation in $arl$ values ins observed. \\[1 pt]
\item For EWMA control chart we observe that the choice of $\lambda$ and $\L$ values play a significant role.Though standard $\L$ may be taken to be 3, but it is very difficult to choose an optimal $\lambda$ uniformly for all choices of depth functions. However we have observed that $\lambda$ in the range of 0.20-0.40 gives better ARL performance compared to others. \\[1 pt]
\item We were not able to get good fits for  the $\psi^2$ and $\tau^2$ statistics data in most cases,with any standard distributions over the entire support. The gamma distribution fits the data well for except for the high end values, which leads to lack of fit. Due to the lack of any standard distribution fit,we had to resort to Bootstrapping. But in practical scenarios, one may still use the gamma distribution in finding the cut-off as the Bootstrap is computationally difficult. It will not be a very bad approximation and will save time. We present the gamma distribution fits to the empirical distribution of $\tau^2$ and $\psi^2$ to assert this point. \\[1 pt]
\end{itemize}

{}

\newpage

\begin{sidewaystable}
\caption{Comparative ARL performance of Shewhart chart with modifications }
\label{table:arluniv}
\begin{tabular}{|p{1.5cm}|r|r|r|r|r|r|r|r|r|r|r|r|r|r|r|r|r|r|r|r|r|}
\hline
Subgroup Size & \multicolumn{3}{|c|}{10} & \multicolumn{3}{|c|}{11} & \multicolumn{3}{|c|}{12} & \multicolumn{3}{|c|}{13} & \multicolumn{3}{|c|}{14} & \multicolumn{3}{|c|}{15} & \multicolumn{3}{|c|}{20}\\
\hline
Shift & 10\% & 20\% & Mean & 10\% & 20\% & Mean & 10\% & 20\% & Mean & 10\% & 20\% & Mean & 10\% & 20\% & Mean & 10\% & 20\% & Mean & 10\% & 20\% & Mean\\
\hline
0 & 162 & 154 & 185  & 190 & 192 & 183  & 161 & 158 & 178  & 163 & 158 & 163  & 185 & 181 & 178  & 151 & 146 & 144  & 146 & 130 & 183 \\
\hline
0.25 & 36 & 45 & 50  & 77 & 46 & 104  & 75 & 34.69 & 82 & 13.6 & 50 & 132 & 30 & 64 & 58 & 54 & 48  & 53 & 25 & 50.66 & 47.6 \\
\hline
0.5 & 8.06 & 9.72 & 9.7  & 12.6 & 9.18 & 15  & 19 & 12.4 & 6.65 & 11 & 3.73 & 8.15 & 15 & 6 & 8.95 & 8.1 & 8.4 & 6.98 & 6.2 & 4.2 & 5.77\\
\hline
0.75 & 2.86 & 3.33 & 3.15 & 3.63 & 3.07 & 4 & 4.88 & 3.53 &2.3  & 3.18 & 1.7 & 2.51 & 3.6 & 2.2 & 2.57 & 2.4 & 2.54 & 2.17 & 1.9 & 1.6 & 1.75\\
\hline
1 & 1.54 & 1.7 & 1.6  & 1.71 & 1.59 & 1.79 & 2.02 & 1.67 & 1.32 & 1.52 & 1.17 & 1.35 & 1.58 & 1.3 &1.34  & 1.3 & 1.36 & 1.23 & 1.1 & 1.1 & 1.1\\
\hline
Outlier &  &  &  &  &  &  &  &  &  &  &  &  &  &  &  &  &  &  &  &  & \\
\hline
1 & 107 & 106  & 25.7  & 145 & 106  & 20.8 & 106 &  107  & 13.9  & 107 & 100  &  60  & 107 & 113  & 37.8 & 128 & 143  &  44.1 & 180 & 139  & 25.1 \\
\hline
2 & 112  & 118 &  12.9  & 99 & 139 & 24.8 & 156 & 112 & 11.7  & 109  & 115 &  23.9 & 100 & 112 & 1.68  & 102  & 120 & 13.5 & 117  & 155 &  24.2 \\
\hline
\end{tabular}
\end{sidewaystable}

\newpage

\begin{sidewaystable}
\caption{Comparative ARL performance of EWMA chart with modifications }
\label{table:arlewma}
\begin{tabular}{|r|r|r|r|r|r|r|r|r|r|r|r|r|}
\hline
$\lambda$ & \multicolumn{4}{|c|}{0.2} & \multicolumn{4}{|c|}{0.25} & \multicolumn{4}{|c|}{0.4}\\
\hline
 & \multicolumn{2}{|c|}{10\% trimming} & \multicolumn{2}{|c|}{Mean} & \multicolumn{2}{|c|}{10\% trimming} & \multicolumn{2}{|c|}{Mean} & \multicolumn{2}{|c|}{10\% trimming} & \multicolumn{2}{|c|}{Mean}\\
\hline
 & ARL & sd(ARL) & ARL & sd(ARL) & ARL & sd(ARL) & ARL & sd(ARL) & ARL & sd(ARL) & ARL & sd(ARL)\\
\hline
Case &  &  &  &  &  &  &  &  &  &  &  & \\
\hline
No Shift & 191.67 & 95.11 & 145 & 67.54 & 132.89 & 33.86 & 161.7 & 39.73 & 181.67 & 46.51 & 157.22 & 37.65\\
\hline
.1 $\sigma$ shift & 14.98 & 5.43 & 23.84 & 7.9 & 20.17 & 3.95 & 39.6 & 22.36 & 32.6 & 11.52 & 66.56 & 18.56\\
\hline
.25 $\sigma$ shift & 1.623 & 0.17 & 1.641 & 0.1003 & 2.082 & 0.194 & 2.063 & 0.132 & 3.995 & 0.482 & 4.984 & 0.681\\
\hline
.5 $\sigma$ shift & 1 & 0.0002 & 0.99 & 0 & 1.0302 & 0.0038 & 1 & 0.015 & 1095 & 0.022 & 1.083 & 0.0119\\
\hline
Outlier & 116.7 & 66.71 & 1.194 & 0.058 & 151.83 & 43.05 & 1.398 & 0.0757 & 129.96 & 33.35 & 2.686 & 0.234\\
\hline
\end{tabular}
\end{sidewaystable}

\begin{sidewaystable}
\caption{Comparative ARL performance of Multivariate $T^2$ chart with modifications }
\label{table:arlmulti}
\begin{tabular}{|r|r|r|r|r|r|r|r|r|r|r|}
\hline
Depth fn & \multicolumn{2}{|c|}{Liu} & \multicolumn{2}{|c|}{Oja} & \multicolumn{2}{|c|}{Spatial} & \multicolumn{2}{|c|}{Tukey} & \multicolumn{2}{|c|}{Mean}\\
\hline
 & ARL & sd(ARL) & ARL & sd(ARL) & ARL & sd(ARL) & ARL & sd(ARL) & ARL & sd(ARL)\\
\hline
Case &  &  &  &  &  &  &  &  &  & \\
\hline
No shift & 158 & 61 & 156 & 59 & 131 & 54 & 134 & 47 & 161 & 72\\
\hline
(.5,.5) Shift & 4.9 & 0.215 & 5.15 & 1.43 & 4.83 & 0.76 & 3.95 & 1.68 & 3.23 & 0.147\\
\hline
(1,1) Shift & 1.02 & 0.074 & 1.42 & 0.078 & 1.18 & 0.094 & 1.08 & 0.56 & 1.03 & 0.005\\
\hline
\multicolumn{3}{|c|}{1 Outlier with mean} &  &  &  &  &  &  &  & \\
\hline
(5,5) & 107 & 35 & 86 & 31 & 147 & 68 & 111 & 38 & 17.26 & 2.16\\
\hline
(5,0) Shift & 135 & 74 & 101 & 32 & 106 & 75 & 119 & 47 & 34 & 3.88\\
\hline
(0,5) Shift & 107 & 48 & 95 & 33 & 137 & 89 & 103 & 33 & 32.27 & 7.45\\
\hline
\end{tabular}
\end{sidewaystable}

\newpage

\begin{sidewaystable}
\caption{Comparative ARL  performance for MEWMA chart with modifications }
\label{table:arlmultimulti}
\begin{tabular}{|p{1.4 cm}|r|r|r|r|r|r|r|r|r|r|r|}
\hline
Depth fn &  & \multicolumn{2}{|c|}{Liu} & \multicolumn{2}{|c|}{Oja} & \multicolumn{2}{|c|}{Spatial} & \multicolumn{2}{|c|}{Tukey} & \multicolumn{2}{|c|}{Mean}\\
\hline
 &  & ARL & sd(ARL) & ARL & sd(ARL) & ARL & sd(ARL) & ARL & sd(ARL) & ARL & sd(ARL)\\
\hline
Case & $\lambda$ &  &  &  &  &  &  &  &  &  & \\
\hline
\multirow{3}{*}{0 shift} & 0.2 & 183.3 & 99.47 & 95 & 22.64 & 159 & 88 & 130 & 21 & 179 & 84.4\\
\cline{2-12}
 & 0.25 & 220.9 & 169 & 131 & 73 & 111 & 61.74 & 150 & 54 & 161 & 68.6\\
\cline{2-12}
 & 0.4 & 112.5 & 22.84 & 175 & 85 & 188 & 104 & 156 & 57 & 183 & 92\\
\hline
\multirow{3}{*}{(.1,.1) } & 0.2 &  &  & 17.7 & 4.82 & 14.5 & 5.1 & 15.7 & 3.9 & 16.29 & 4.23\\
\cline{2-12}
 & 0.25 & 10.54 & 2.34 & 19.17 & 6.27 & 14.72 & 3.28 & 18.95 & 4.48 & 10.63 & 1.54\\
\cline{2-12}
 & 0.4 & 16.4 & 3.63 & 22.56 & 4.81 & 22.77 & 7.95 & 24 & 5.13 & 20.36 & 3.07\\
\hline
\multirow{3}{*}{(.25,.25) } & 0.2 & 1.35 & 0.043 & 1.26 & 0.023 & 1.34 & 0.038  & 1.49 & 0.1 & 1.14 & 0.023\\
\cline{2-12}
 & 0.25 & 1.45 & 0.05 & 1.8 & 0.112 & 1.74  & 0.14 & 1.7 & 0.16 & 1.9 & 0.125\\
\cline{2-12}
 & 0.4 & 2.4 & 0.169 & 3.65 & 0.469 & 2.55 & 0.48  & 3.01 & 0.257 & 2.67 & 0.169\\
\hline
\multicolumn{3}{|c|}{1 Outlier with mean} &  &  &  &  &  &  &  &  & \\
\hline
\multirow{3}{*}{(3,3)} & 0.2 & 83.27 & 46 & 19.14 & 4.98 & 85.41 & 32.18 & 9.12 & 1.12 & 1.43 & 0.063\\
\cline{2-12}
 & 0.25 & 92 & 10.9 & 42 & 9.99 & 89.65 & 40.8 & 129.5 & 114.8 & 5.88 & 0.547\\
\cline{2-12}
 & 0.4 & 84 & 28.85 & 400 & 367 & 102 & 57 & 127.7 & 92 & 10.1 & 1.137\\
\hline
\end{tabular}
\end{sidewaystable}

\newpage 

\begin{table}
\caption{Comparative ARL performance of our method with Liu's method and \\ Chenouri \emph{et al}'s method}
\label{table:liumcd}
\begin{tabular}{|r|r|r|r|r|r|r|}
\hline
Method & \multicolumn{2}{|c|}{Our method} & \multicolumn{2}{|c|}{Liu's method} & \multicolumn{2}{|c|}{MCD based method}\\
\hline
 & ARL & sd(ARL) & ARL & sd(ARL) & ARL & sd(ARL) \\
\hline
Case &  &  &  &  &  &   \\
\hline
No shift & 199.34 & 91 & 159.3 & 32.36 & 126.4 & 43.76\\
\hline
.25 Shift & 24.9 & 6.3 & 13.4 & 2.6 & 60.04 & 34.4 \\
\hline
0.5 Shift & 11.02 & 1.28 & 9.88 & 1.44 & 29.223 & 10.41 \\
\hline
1 Shift & 2.02 & 0.074 & 1.2 & 0.034 & 11.52 & 1.927 \\
\hline
\multicolumn{4}{|c|}{1 Outlier from $N(3,3)$ } &  &  &  \\
\hline
 & 104.78 & 35.12 & 16.68 & 4.27 & 94.33 & 22.62\\
\hline
\multicolumn{4}{|c|}{2 Outliers from $N(3,3)$ } &  &  & \\
\hline
 & 94.66 & 28.34 & 19.27 & 5.48 & 78.44 & 18.16\\
\hline
\end{tabular}
\end{table}

\newpage

\begin{thebibliography}{}

\bibitem [Shewhart(1931)]{shewhart:31}
Shewhart, Walter A.(1931).
\newblock Economic Control of Quality of Manufactured Product, New York: D. Van Nostrand Company, Inc.

\bibitem[Azzalini(2005)]{azza:05}
Azzalini, A. (2005).
\newblock The skew-normal distribution and related multivariate families.
\newblock \emph{Scandinavian Journal of Statistics}, {32}, 159--188.

\bibitem[Hotelling(1931)]{hotelling:31}
Hotelling H (1931)
\newblock The Generalization of Student's Ratio 
\newblock \emph{Annals of Mathematical Statistics }

\bibitem [Alfaro(2008)]{alfaro:08}
Alfaro, Jos�e Luis and Ortega, Juan Fco. (2008).
\newblock A Robust Alternative to Hotelling�s $T^2$ Control Chart Using Trimmed Estimators.
\newblock \emph{Quality Reliability Engineering International}, {24},  601-�611.

\bibitem [Baguio(2008)]{baguio:08}
Baguio, C.B. (2008).
\newblock Trimmed Mean as an Adaptive Robust Estimator of a Location Parameter for Weibull Distribution.
\newblock \emph{World Academy of Science, Engineering and Technology}, 681--686.

\bibitem [Barnett(1976)]{barnett:76}
Barnett, V.(1976).
\newblock The Ordering of Multivariate Data.
\newblock \emph{Journal of the Royal Statistical Society. Series A (General)}, {139}, 318--355.

\bibitem [Chambers(1983)]{chambers:83}
Chambers, J. M., Cleveland, W. S., Kleiner, B. M., and Tukey Paul A.(1983).
\newblock Graphical Methods for Data Analysis, Chapman and Hall, New York, 1983.

\bibitem [Chaudhuri(1996)]{chaudhuri:96}
Chaudhuri, P.(1996).
\newblock On a Geometric Notion of Quantiles for Multivariate Data.
\newblock \emph{Journal of the American Statistical Association}, {91}, 862--872.

\bibitem [Crowder (1987)]{crowder:87}
Crowder and Stephen, V.(1987).
\newblock A Simple Method for Studying Run-Length Distributions of Exponentially Weighted Moving Average Charts.
\newblock \emph{Technometrics}, {29}, 401--407.

\bibitem [Gupta (2008)]{gupta:2008}
Gupta, A. and Sengupta, S.(2008).
\newblock Online Control Charts for Process Averages Based on Repeated Median Filters.
\newblock \emph{Communications in Statistics - Simulation and Computation}, {37}, 178 - 202.

\bibitem [Hampel(2011)]{hampel:01}
Hampel, F.(2001).
\newblock Robust statistics: A brief introduction and overview, Invited talk in the Symposium �Robust Statistics and Fuzzy Techniques in Geodesy and GIS� held in ETH Zurich, March 12-16, 2001.

\bibitem [Huber(1972)]{huber:72}
Huber, P.J.(1972).
\newblock The 1972 Wald Lecture:Robust Statistics: A Review.
\newblock \emph{Annals of Mathematical Statistics}, {43}, 1041--1067.

\bibitem [Huber(1980)]{huber:80}
Huber, P.J.(1980).
\newblock Robust Statistical Procedure, 2nd edition, CBMS-NSF REGIONAL CONFERENCE SERIES IN APPLIED MATHEMATICS, Issue 68.

\bibitem [Huber(2002)]{huber:02}
Huber, P.J.(2002).
\newblock John Tukey's Contributions to Robust Statistics.
\newblock \emph{The Annals of Statistics}, {30}, 1640--1648.

\bibitem [Hunter(1986)]{hunter:86}
Hunter, J. S. (1986).
\newblock The exponentially weighted moving average,
\newblock \emph{Journal of Quality Technology}, {18}, 203--210.

\bibitem [Igle(1986)]{igle:86}
Iglewicz, B. and Langenberg, P.(1986).
\newblock Trimmed mean $\bar{X}$ and R charts.
\newblock \emph{Journal of Quality Technology}, {18}, 152--161.

\bibitem [Kochanski(2005)]{kochanski:05}
Kochanski, G. (2005).
\newblock Brute Force as Statistical Tool.

\bibitem [Liu (1988)]{liu:88}
Liu, R. Y. (1988).
\newblock On a notion of simplicial depth,
\newblock \emph{Proceedings of the National Academy of Science USA}, {85}, 1732--1734.

\bibitem [Liu(2006)]{liu:06}
Liu, R. Y., Serfling, Robert J. and Souvaine, Diane L.(2006).
\newblock Data depth: robust multivariate analysis, computational geometry, and Applications.
\newblock DIMACS Series in Discrete Mathematics and Theoretical Computer Science, {72}, American Mathematical Society.

\bibitem[Liu (1995)]{liu:95}
Liu R.Y. (1995)
\newblock Control charts for Multivariate Processes
\newblock \emph{Journal of American Statistical Association}, Vol. 90 No. 432, 1380-1387

\bibitem [Lowry(92)]{lowry:92}
Lowry, Cynthia A., Woodall, William H., Champ, Charles W.and Rigdon Steven E.(1992).
\newblock Multivariate Exponentially Weighted Moving Average Control Chart,
\newblock \emph{Technometrics}, {34}, 46--53.

\bibitem [Lucas(1990)]{lucas:90} 
Lucas, James, M. and Saccucci, Michael S.(1990).
\newblock Exponentially weighted moving average control schemes: properties and enhancements.
\newblock \emph{Technometrics}, {32}, 1--29.

\bibitem [Mont(1996)]{mont:06} 
Montgomery, D.C.(2008). Introduction to Statistical Quality Control, 6th Edition, WILEY.

\bibitem [Oja(1983)]{oja:83} 
Oja, H. (1983).
\newblock Descriptive statistics for multivariate distributions.
\newblock \emph{Statistics \& Probability Letters}, {1}, 327--332.

\bibitem [Roberts(1959)]{roberts:59} 
Roberts, S.W. (1959).
\newblock Control Chart Tests Based on Geometric Moving Averages.
\newblock \emph{Technometrics}, {1}, 239--250.

\bibitem [Stigler(1931)]{stig:73}
 Stigler, S. M.(1973).
\newblock The Asymptotic Distribution of Trimmed Mean,
\newblock \emph{The Annals of Statistics}, 472--477.

\bibitem [Tukey(1963)]{tukey:63}
Tukey, John W. and McLaughin, Donald H.(1963).
\newblock Less vulnerable confidence and significance procedures for local based on a single sample(trimming/winsorizing I).
\newblock \emph{Sankhya}, 331-352.

\bibitem[Chenouri(2007)]{chenouri:07}
Chenouri, Shoja'eddin, Variyath, Asokan M. and Steiner, Stefan H. (2007).
\newblock A Multivariate Robust Control Chart for Individual Observations.
\newblock \emph{\url{http://sas.uwaterloo.ca/stats_navigation/techreports/07WorkingPapers/2007-07.pdf}}.

\end{thebibliography}
\end{document}